\documentclass[12pt]{elsarticle}
\usepackage{amsthm,amssymb}
\usepackage{graphicx}
\usepackage{braket}
\usepackage[breaklinks=true,colorlinks=true,linkcolor=black,urlcolor=blue,citecolor=blue]{hyperref}

\newcommand{\ignore}[1]{}
\newcommand{\HH}{\mathcal{H}}
\newcommand{\binom}[2]{{{#1} \choose {#2}}}

\begin{document}

\title{A graph with fractional revival}

\author[crm]{Pierre-Antoine Bernard}
\ead{BernardPierreAntoine@outlook.com}

\author[york]{Ada Chan}
\ead{ssachan@yorku.ca}

\author[crm]{\'{E}rika Loranger}
\ead{erikaloranger@yahoo.ca}

\author[cu]{Christino Tamon}
\ead{ctamon@clarkson.edu  }

\author[crm]{Luc Vinet\corref{cor1}}
\ead{luc.vinet@umontreal.ca}

\address[crm]{Centre de Recherches Math\'{e}matiques, Universit\'{e} de Montr\'{e}al, 
	C.P, 6128, Montr\'{e}al, QC, Canada, H3T 3J7}

\address[york]{Department of Mathematics and Statistics, York University, 
	Toronto, ON, Canada, M3J 1P3}

\address[cu]{Department of Computer Science, Clarkson University,
	Postdam, NY, USA, 13699-5815}

\cortext[cor1]{Corresponding author}

\begin{abstract}
An example of a graph that admits balanced fractional revival between antipodes is presented. 
It is obtained by establishing the correspondence between the quantum walk on a hypercube where 
the opposite vertices across the diagonals of each face are connected and, the coherent transport of 
single excitations in the extension of the Krawtchouk spin chain with next-to-nearest neighbour interactions.

\end{abstract}

\maketitle

\section{INTRODUCTION}

This article provides an example of a graph that exhibits balanced fractional revival (FR) at two sites. 
FR \cite{r04} is observed during a quantum walk process when an initially localized state evolves so that 
after some time it has non-vanishing probability amplitude uniquely at a number of isolated places. 
FR is said to be balanced if these probabilities are equal at each site. This phenomenon has been shown 
to occur in spin chains \cite{bcb15, gvz16}. In these systems, a state having at first a single excitation 
at one end is found, after finite time, with non-zero amplitudes for this spin up, only at both ends. 
Balanced FR can thus serve as a mechanism to generate maximally entangled states.
\medskip

A special case of fractional revival is when the revival takes place at a single site (with probability one). 
If this site is the one where the spin up was initially located, perfect return occurs. Otherwise, if it is a 
different site (for instance the opposite end of the chain), one speaks of perfect state transfer (PST). 
PST in spin chains is attracting much interest \cite{b07, k10, nj14} especially for the design of quantum 
wires that would require a minimum of external control. PST can also be realized in photonic lattices, 
i.e. in arrays of optical waveguides \cite{perez-leija13, chapman16}. It has been shown that non-uniform 
couplings are required in these devices for PST beyond a few sites \cite{cddekl05}. Favored models are 
those where the couplings between the nearest neighbours are related to the recurrence coefficients of 
the Krawtchouk polynomials; they are often referred to by that name for that reason \cite{acde04}.
\medskip

These advances have prompted the study of PST in spin networks; here the one excitation Hamiltonian 
is provided by the adjacency matrix of the underlying graph. PST on graphs has been much analysed and 
the reader is invited to read \cite{g12} and \cite{kt11}  for reviews. Little consideration has been 
given to FR on graphs however. A significant observation that has been made \cite{cddekl05,cfg02} is 
that PST between antipodes of the one-link hypercube is tantamount to end-to-end perfect transport in 
the Krawtchouk spin chain. This is established by showing that the quantum walk on the graph projects 
onto the one on the line or $P_{N}$, with appropriate weights for the $N$ links. PST occurs in the 
Krawtchouk chain with nearest-neighbour (NN) couplings; it is however easy to check that FR is not 
possible in this model. In view of the preceding point, the same goes for the hypercube, namely this 
graph will not support FR.
\medskip

As it turns out, an extension of the NN-Krawtchouk model that includes interactions between next-to-nearest 
neighbours (NNN) has been designed recently \cite{cvz17} and found to admit balanced FR under certain 
conditions. We shall use these results here, to find a graph with FR, by showing that the quantum walk between antipodal 
points on that graph projects equivalently to the NNN one-excitation dynamics of the chain known to have FR. 
This graph will be identified as a hypercube where the opposite vertices across the diagonals on each face 
are connected and where all these links have the same weight relative to the one attributed to the edges 
of the hypercube.
\medskip

The remainder of the paper will proceed as follow. We shall first remind the reader of the NNN extension of 
the Krawtchouk spin chain and of its PST and FR properties. We shall follow by reviewing basic features of 
Hamming graphs and their adjacency matrices. Their relation to the Krawtchouk polynomials will be brought up. 
We shall further recall how the connection between quantum walks on the hypercube and on the one-dimensional 
graph with Krawtchouk weights is made by restricting the adjacency matrix to an appropriate ``column" subspace. 
We shall then determine the graph whose restriction to that subspace yields the NNN one-excitation dynamics. 
Concluding remarks will be offered and the paper will end with an Appendix where the occurrence of FR on the graph 
is verified directly.

\section{FRACTIONAL REVIVAL IN THE KRAWTCHOUK SPIN CHAIN WITH NEXT-TO-NEAREST NEIGHBOUR INTERACTIONS}

We shall consider a spin chain with the following Hamiltonian of type XX on $(\mathbb{C})^{\otimes N}$ 
where $N$ spins interact with their nearest and next-to-nearest neighbours:
\begin{equation}
H = 
\frac{1}{2} \sum_{n = 1}^N  J_{n}^{(1)}({\sigma}_{n}^{x}{\sigma}_{n+1}^{x} 
	+ {\sigma}_{n}^{y}{\sigma}_{n+1}^{y}) + J_{n}^{(2)}({\sigma}_{n}^{x}{\sigma}_{n+2}^{x} 
	+ {\sigma}_{n}^{y}{\sigma}_{n+2}^{y}) + B_n \sigma^{z}_{n}
\end{equation}
where $J_N^{(1)} = 0$ and $J_{N}^{(2)} = J_{N-1}^{(2)} = 0$. 
As usual, ${\sigma}_{n}^{x}$,  ${\sigma}_{n}^{y}$ and  ${\sigma}_{n}^{z}$ stand for the Pauli matrices 
with the index $n$ indicating on which of the $\mathbb{C}$ factors they act. 
\medskip

The coupling constants are built from the recurrence coefficients of the Krawtchouk polynomials.
(See (\ref{eqn:krawtchouk_recurrence}) below.) 
Let
\begin{equation} \label{eqn:jn_coupling}
J_n = \frac{1}{2} \sqrt{n(N-n)}.
\end{equation}
The nearest-neighbour (NN) couplings are taken to be
\begin{equation} \label{eqn:nn_jn_coupling}
J_n^{(1)} = \beta J_n
\end{equation}
with $\beta$ an energy scale parameter. The next-to-nearest neighbour (NNN) couplings are chosen as 
\begin{equation} \label{eqn:nnn_jn_coupling}
J_n^{(2)} = \alpha J_n J_{n+1}
\end{equation}
with $\alpha$ another parameter. Finally, the Zeeman terms will be specified by 
\begin{equation}
B_n = \alpha (J_n^2 + J_{n-1}^2).
\end{equation}
Observe that when $\alpha = 0$, $J_n^{(2)} = B_n = 0$ and the NN Krawtchouk model with no magnetic field 
is recovered. Note also that slight changes have been made with respect to \cite{cvz17}: 
we are here taking $N$ (instead of $N+1$) to be the total number of sites and have modified 
the range of $n$ accordingly. 

\medskip

Owing to rotational symmetry around the $z$-direction, $H$ preserves the total number of spins that are 
up along the chain. It will suffice here to consider chain states that have only one excitation or spin up. 
A natural basis for that subspace is given by the vectors
\begin{equation}
\ket{n} = (0, 0, ..., 0, 1, ... 0)^T 
\ \hspace{1in} \ n = 1, ..., N 
\end{equation}
with the only $1$ in the $n$th position. This vector is associated to a single spin up at the $n$th site. 
The action of $H$ on those state is 
\begin{equation}
H\ket{n} = J_{n}^{(2)}\ket{n+2} + J_{n}^{(1)}\ket{n+1} 
	+ B_{n}\ket{n} + J_{n-1}^{(1)}\ket{n-1} + J_{n-2}^{(2)}\ket{n-2}.
\end{equation}
We remark that if we define the matrix $J$ by
\begin{equation} \label{eqn:nn_coupling}
J\ket{n} =  J_{n}\ket{n+1}  + J_{n-1}\ket{n-1} 
\end{equation}
we have
\begin{equation} \label{eqn:nnn_coupling}
H\ket{n} = (\alpha J^2 + \beta J)\ket{n}.
\end{equation}
Fractional revival (FR) at two sites occurs if there is a time $\tau_{FR}$ such that 
\begin{equation}
e^{-i\tau_{FR}H}\ket{n} = \mu \ket{0} + \nu\ket{N}
\end{equation}
with $\mu$,$\nu \in \mathbb{C}$  such that $|\mu|^2+|\nu|^2 = 1$. In other words, FR takes place if 
the dynamics allows to evolve the state with a spin up localized at site $1$ into a state that is 
``revived'' at both ends of the chain. FR is balanced when $|\mu|=|\nu| = \frac{1}{\sqrt{2}}$ in 
which case, a maximally entangled state has been generated at time $\tau_{FR}$. 
The special case of FR that happens when $\mu = 0$ is referred to as perfect state transfer (PST) since 
the spin up at site $1$ is then transported at site $N$ with probability one after a time we will denote 
$\tau_{PST}$. The NN Krawtchouk chain ($\alpha = 0$) is well known to admit PST at 
$\tau_{PST} = \frac{\pi}{\beta}$.

\medskip

The coherent transport of single excitation along the NNN Krawtchouk spin has been studied in \cite{cvz17}. 
In summary, the findings are as follows. It is first noted that FR can not be found in the NN situation 
when $\alpha = 0$. When $\alpha \neq 0$, balanced FR can happen. Apart from an overall phase factor, 
$\mu$ would be real, $\nu$ pure imaginary and $|\mu|=|\nu| = \frac{1}{\sqrt{2}}$. There are conditions 
on $\alpha$, $\beta$ and $N$.
\begin{enumerate}
\item[i.] {\em Case} $\beta \neq 0$: \\ 
The ratio $\frac{\alpha}{\beta}$ must be a rational number:
\begin{equation}
\frac{\alpha}{\beta} = \frac{p}{q}
\end{equation}
where $p$ and $q$ are coprime integers. FR will happen if in addition,
\begin{itemize}
\item $p$ is odd
\item $q$ and $N$ have different parities. 
\end{itemize}
The time at which balanced FR will then occur first is 
\begin{equation}
\tau_{FR} = \frac{\pi q}{2 \beta}.
\end{equation}
PST will be found in these circumstances at double the FR time: $\tau_{PST} = 2\tau_{FR}$. 
PST is also possible at $\tau_{PST} = \frac{\pi q}{\beta}$ if $p$ is even and $q$ odd even though 
FR can not be realized in this case.

\item[ii.] {\em Case} $\beta = 0$: \\ 
FR and PST as well, are possible only when $N$ is odd. The minimal times are for FR, 
\begin{equation}
\tau_{FR} = \frac{\pi}{2 \alpha}
\end{equation}
and for PST, $\tau_{PST} = 2\tau_{FR}$.

\end{enumerate}

\section{ELEMENTS OF THE BINARY HAMMING SCHEME} \label{section:binary_hamming}

We shall now review properties of certain Hamming graphs that will be used to obtain a lift to a graph 
with FR, of the single excitation dynamics of the spin chain with NNN couplings that we have described 
in the last section. 

\medskip 

Recall that a graph $G = (V,E)$ consists of a set $V$ of vertices and of a set $E$ of edges that are 
two-element subsets of $V$. Edges might be assigned weights. With $|V|$ the cardinality of $V$, 
the adjacency matrix of a graph $G$ is the $|V| \times |V|$ matrix whose $A_{xy}$ entry for $x,y \in V$ 
is equal to the number of edges between the vertices $x$ and $y$. 

\medskip

We shall focus on graphs in the binary Hamming scheme \cite{bcn89} $\HH(M,2)$. In this case, 
$V = \{0,1\}^{M}$, i.e., vertices are labelled by sequences of $M$-tuples of $0$s and $1$s. 
The Hamming distance $d(x,y)$ between the vertices $x$ and $y$ is the number of positions where the 
$M$-tuples $x$ and $y$ differ. We shall denote by $G_i, i = 0, 1,\ldots, M$, the graph whose edges are 
connecting all pairs of vertices with Hamming distance $i$. $G_1$ is the $M$-dimensional hypercube: 
the edges connect the $M$-tuples which differ in exactly one place. $G_2$ has the same vertices as $G_1$, 
but the edges are between opposite vertices of each face of the hypercube. 

\medskip

We shall denote by $A_i$ the adjacency matrix of the graph $G_i$. The intersection numbers 
$p_{ij}^{k} = p_{ji}^{k}$ count how many $z \in V$ there are such that $d(x,z) = i$ and 
$d(y,z) = j$ if $d(x,y) = k$. It follows that the matrices $A_0 = I, A_1, \ldots, A_M$ verify the relations
\begin{equation}
A_i A_j = \sum_{k = 0}^{M} p_{ij}^{k} A_k 
\ \hspace{1in} \ 
i,j \in \{1, 2, \ldots, M\}
\end{equation}
of the Abelian Bose-Mesner algebra \cite{bcn89,bi84,b90}.

\medskip

It is easily seen \cite{b90,s01} that $p_{i1}^{k}$ is a tridiagonal matrix. Indeed, 
\begin{equation} \label{eqn:krein}
A_i A_1 = c_{i+1}A_{i+1} + b_{i-1}A_{i-1} 
\ \hspace{1in} \ 
0 \le i \le M
\end{equation}
with
\begin{equation} \label{eqn:krein2}
c_{i+1} = i + 1 
\ \hspace{1in} \ 
b_{i-1} = M - i + 1
\end{equation}
where $c_{M+1} = b_{-1} = 0$.
This is found as follows. The number $c_{i+1} = p_{i1}^{i+1} $ counts how many $z$ there are 
such that $d(x,z) = i$ and $d(y,z) = 1$ if $d(x,y) = i+1$. 
To find this number, take $x = (0, 0, \ldots, 0)$ and $y = (1, 1, \ldots, 1, 0, \ldots, 0)$ 
with $i+1$ $1$s, this ensures that $d(x,y) = i+1$.
It is seen that $z$ must be obtained from $y$ by changing one of the $1$s to a $0$ and there are 
$i+1$ ways of doing this. Similarly, $b_{i-1} = p_{i1}^{i-1} $ counts the number of $z$ such that 
$d(x,z) = i$ and $d(y,z) = 1$ if $d(x,y) = i-1$. Picking $x = (0, \ldots, 0)$ and 
$y = (1, \ldots, 1, 0, \ldots, 0)$ with $i - 1$ $1$s in this case, we observe that $z$ must be obtained 
from $y$ by converting one of the $0$s to a $1$ and there are $M - i + 1$ possibilities. 
Finally, it is readily noted that there are no $z$ such that $d(x,z) = i$ and $d(y,z) = 1$ 
when $d(x,y) = i$ and hence $p_{i1}^{i} = 0$.

\medskip

A particular case of (\ref{eqn:krein}) and (\ref{eqn:krein2}) is
\begin{equation} \label{eqn:nn_squared}
A_1^{2} = 2A_{2} - M \cdot I
\end{equation}
This relation will be used in the next section. More generally, it follows from (\ref{eqn:krein}) 
that $A_i = p_i(A_1)$ with $p_i(A_1)$ a polynomial of degree $i$ in $A_1$. By considering the $M$ 
eigenvalues $\lambda_s$, $s=1,\ldots,M$, of $A_1$ which are distinct, we have a real-valued function 
which is expressed as follows
\begin{equation}
p_i(\lambda_s) = K_i(s; 2, M)
\end{equation}
with $\lambda_{s} = M-2s$ and initial conditions $K_{-1} = 0$, $K_{0} = 1$,
in terms of the Krawtchouk orthogonal polynomials $K_n(x; q, M)$ defined by the recurrence relation
\begin{equation} \label{eqn:krawtchouk_recurrence}
(n+1)K_{n+1}(x) = [(M-n)(q-1) + n - qx]K_n(x) - (q-1)(M-n+1)K_{n-1}(x).
\end{equation}
The appearance of these orthogonal polynomials in both the spin chain and the Hamming scheme 
$\HH(M,2)$ is already suggestive of a connection that we shall make explicit in what follows.

\section{PROJECTING A WEIGHTED HAMMING GRAPH ONTO THE SPIN CHAIN}

It has been shown in \cite{cddekl05} that the NN Krawtchouk spin chain can be obtained by projecting 
quantum walks on the one-link hypercube onto a ``column'' subspace. We shall review how this is done 
and then build upon this result to identify the graph that will exhibit the same dynamics as that of 
the NNN Krawtchouk chain. This will yield in particular the conditions for that graph to exhibit FR. 

\medskip

To establish the correspondence, we shall take the parameter $M$ of 
Section \ref{section:binary_hamming} to be $M = N - 1$. To the vertices 
$x \in V = \{0,1\}^{N - 1} $, we shall associate orthonormalized vectors 
$\ket{x} \in \mathbb{C}^{|V|}$, with $|V| = 2^{N-1}$, such that
\begin{equation}
\braket{x|y} =
	\left\{\begin{array}{ll}
    1 & \mbox{ if $d(x,y) = 0$ } \\
    	0 & \mbox{ otherwise }
    \end{array}\right.
\end{equation}
%Il faudrait ajouter l'équation comme une matrice ici ?
for $x, y \in V$. In this framework the entries of the adjacency matrix $A$ are given by 
$A_{xy} = \bra{x}A\ket{y}$. Let $(0) \equiv (0, 0, \ldots, 0)$ denote a corner and organize 
$V$ as a set of $N$ columns $V_n$, $n = 1,\ldots, N$, defined by 
\begin{equation}
V_n = \{x \in V \ : \ d(0,x) = n - 1\}.
\end{equation}
In other words, the vertices of the $n$th column are $n - 1$ edges away from the corner $(0)$. 
The number $k_n$ of vertices in the column $V_{n}$ is therefore given by $k_n = {{N - 1} \choose {n - 1}}$. 
Let us then label by $V_{n,m}$, $m = 1, \ldots, k_n$, the vertices in column $V_{n}$, every one has $n - 1$ $1$s. 
In $G_1$, each $V_{n,m}$ is connected to the $(N-n)$ elements of column $V_{n + 1}$ obtained by 
converting a $0$ of this $V_{n,m}$ to a $1$.

\medskip

Consider now the column space taken to be the linear span of the column vectors given by
\begin{equation} \label{eqn:basis_column}
\ket{col \  n} = \frac{1}{\sqrt{k_n}}\sum_{m = 1}^{k_n}\ket{V_{n,m}} 
\ \hspace{1in} \ n = 1, \ldots, N.
\end{equation}
The key observation, already made in \cite{cfg02}, is that the evolution with the adjacency matrix $A_1$ 
preserves the column space because every vertex in column $V_{n}$ is connected to the same number of vertices 
in column $V_{n + 1}$ and vice-versa. It follows that we can project on column space when considering 
quantum walks on $G_1$ starting at $(0)$

\medskip

Following \cite{cddekl05}, let us compute the matrix elements of $A_1$ in the basis (\ref{eqn:basis_column}) 
of the column subspace:
\begin{eqnarray} \label{eqn:braket_column}
\bra{col \ n+1} A_1 \ket{col \ n} 
	& = & \frac{1}{\sqrt{k_{n + 1}k_n}} 
    	\sum_{m'=1}^{k_{n+1}}\sum_{m=1}^{k_n}\bra{V_{n + 1, m'}}A_1 \ket{V_{n,m}} \\
	& = & \frac{1}{\sqrt{k_{n + 1}k_n}} k_n (N-n) \\
	& = & \sqrt{n(N-n)} = 2 J_n.
\end{eqnarray}
In obtaining (\ref{eqn:braket_column}), one can first pick a vertex in $V_n$, compute the scalar products 
with the $(N - n)$ vertices to which it is linked in $V_{n + 1}$ and then sum over the $k_{n}$ vertices 
in $V_n$. We thus observe that 
\begin{equation} \label{eqn:weighted_nn_coupling}
\bra{col \ n+1} A_1 \ket{col \ n} = 2 \bra{n+1} J \ket{n}
\end{equation}
with $J$ specified in (\ref{eqn:nn_coupling}). 

There is no need to repeat the argument for 
$\bra{col \ n-1} A_1 \ket{col \ n}$ since the outcome is given by symmetry. 
Owing to the fact that $A_{xy} = A_{yx}$, 
\begin{equation}
\bra{col \ n-1} A_1 \ket{col \ n} = \bra{col \ n} A_1 \ket{col \ n - 1} = 2 J_{n-1}.
\end{equation}
We thus reach the conclusion that the hypercube $G_1$ has the same behavior as the NN Krawtchouk 
spin chain and we may infer that PST occurs on the hypercube knowing that this is so along the spin chain.

\medskip

Let us now turn our attention to the NNN model and look for a graph that will project onto it under 
restriction to the same column space. In view of the fact that the 1-excitation Hamiltonian of the NNN 
spin chain is given by $\alpha J^2 + \beta J$ (see (\ref{eqn:nnn_coupling})), it is clear from 
(\ref{eqn:weighted_nn_coupling}) that the matrix 
$H_G = \frac{\alpha}{4}A_1^2 + \frac{\beta}{2}A_1$ will project onto the desired dynamics in one dimension. 
This $H_G$ can be expressed as the adjacency matrix of a weighted graph by recalling the relation between 
$A_1$ and $A_2$, given (\ref{eqn:nn_squared}). This leads to 
\begin{equation}
H_G = \frac{\alpha}{2}A_2 + \frac{\beta}{2}A_1 + \frac{\alpha}{4}(N - 1)I
\end{equation}
for the graph Hamiltonian $H_G$. The constant term is simply an overall energy shift. One can hence conclude 
that the appropriate graph is the union $G_1 \cup G_2$ with the edges of $G_2$ weighted by the relative 
factor $\frac{\alpha}{\beta}$ with respect to those of $G_1$.

\medskip

We shall confirm this proposition directly by determining how $A_2$ projects onto column space. 
By definition, $A_2$ connects vertex vectors associated to $(N-1)$-tuples that are at distance $2$ 
one from an other. Within the column picture, there are two possibilities for this to happen. 
The first is between elements of the columns $V_n$ and $V_{n \pm 2}$ since they are respectively 
at distance $n - 1$ and $n + 1$ or $n - 3$ from $(0)$. The second is between elements of a single 
column $V_n$ that differ in two positions, one such $G_2$ edge would be between the vertices 
$(1, \ldots, 1, 0, \ldots, 0) $ and $(0, 1, \ldots, 1, 0, \ldots, 0) $ each with $n - 1$ $1$s.

\medskip

Let us first consider $\bra{col \ n+2} A_2 \ket{col \ n}$. Take a $V_{n,m}$ in $V_n$. To obtain an 
element in $V_{n+2}$, one needs to replace two $0$s by two $1$ in the $(N-1)$-tuples $V_{n,m}$ which 
has $(n-1)$ $1$s. One thus finds $\frac{1}{2}(N-n)(N-n-1)$ different vertices in $V_{n+2}$ that are 
at distance $2$ from $V_{n,m}$ in $V_n$ and one arrives at the following:
\begin{eqnarray}
\bra{col \ n + 2} A_2 \ket{col \ n} 
	& = & \frac{1}{\sqrt{k_{n + 2}k_n}} \sum_{m'=1}^{k_{n+2}}
		\sum_{m = 1}^{k_n}\bra{V_{n + 2, m'}}A_2 \ket{V_{n,m}} \\
	& = & \frac{1}{\sqrt{2k_{n + 2}k_n}} k_n (N-n)(N-n-1) \\
	& = & \frac{1}{2}\sqrt{n(n+1)(N-n)(N-n-1)}
\end{eqnarray}
using $k_n = {N-1 \choose n-1}$. We thus see that 
\begin{equation} \label{eqn:quotient_nnn_coupling_forward}
\bra{col \ n + 2} A_2 \ket{col \ n} = 2 J_n J_{n+1}
\end{equation}
and by symmetry one has 
\begin{equation} \label{eqn:quotient_nnn_coupling_backward}
\bra{col \ n - 2} A_2 \ket{col \ n} = 2 J_n J_{n-1}.
\end{equation}

\medskip

Let us examine now $\bra{col \ n} A_2 \ket{col \ n}$. Again pick a vertex $V_{n,m}$ in $V_{n}$. 
All vertices obtained from $V_{n,m}$ by replacing one $1$ by one $0$ and one $0$ by one $1$ will 
still be in $V_n$ and at a distance $2$ from $V_{n,m}$. (Those obtained by interchanging $\ell$ $0$s 
and $1$s will again be in $V_n$ but at greater distance $2\ell$ if $\ell > 1$). There are $(n - 1)(N - n)$ 
ways of picking a $1$ and a $0$ to permute them. Hence, 
\begin{eqnarray}
\bra{col \ n} A_2 \ket{col \ n} 
	& = & \frac{1}{k_n} \sum_{m'=1}^{k_{n}}\sum_{m = 1}^{k_n}\bra{V_{n, m'}}A_2 \ket{V_{n,m}} \\
	& = & (n - 1)(N - n).
\end{eqnarray}
Now note that
\begin{eqnarray}
J_n^2 + J_{n-1}^2 
	& = & \frac{1}{4}[ n(N-n)+(n-1)(N-n+1)] \\
	& = & \frac{1}{4}[2(n-1)(N-n)+(N-1)].
\end{eqnarray}
It follows that 
\begin{equation} \label{eqn:quotient_a2_coupling}
\bra{col \ n} \frac{A_2}{2} + \frac{1}{4}(N - 1)I \ket{col \ n} = J_n^2 + J_{n-1}^2.
\end{equation}

\medskip

Equation (\ref{eqn:quotient_nnn_coupling_forward}), 
(\ref{eqn:quotient_nnn_coupling_backward}) 
and (\ref{eqn:quotient_a2_coupling}) 
show not surprisingly that $ \frac{1}{2}A_{2} + \frac{1}{4}(N - 1)I$ has the same matrix 
elements in the column vector basis as $\frac{1}{4}A_1^2$. This confirms that we may take
\begin{equation} \label{eqn:hamiltonian_a12}
\overline{H}_G = \frac{\alpha}{2} A_2 + \frac{\beta}{2} A_1
\end{equation}
as a lift on $V = \{0,1\}^{N-1}$ (up to a constant) of the $1$-excitation Hamiltonian of 
the NNN Krawtchouk chain. From this correspondence, we can claim that $G_1 \cup G_2$ offers an example 
of a graph where FR can take place. This graph corresponds to an hypercube where vertices across the 
two diagonals of each face are connected by an edge in addition to those of the cube. 
The graph Hamiltonian is given by (\ref{eqn:hamiltonian_a12}). It follows that the edges of $G_1$ 
must all have weight $\frac{\beta}{2}$ and those of $G_2$ be given weight $\frac{\alpha}{2}$. 
The conditions and time for FR are the same as those given 
in (\ref{eqn:jn_coupling}), 
(\ref{eqn:nn_jn_coupling}) 
and (\ref{eqn:nnn_jn_coupling}) 
for the NNN Krawtchouk chain. 
If $\beta \neq 0$, we must have $\frac{\alpha}{\beta} = \frac{p}{q}$ 
with $p$ and $q$ coprime integers. 
Furthermore, for FR to happen $p$ must be odd and $q$ and $N$ must have opposite parities. 
Under those conditions, FR will happen at time 
$\tau_{FR} = \frac{\pi q}{2\beta}$. 
If $\beta = 0$, $\overline{H}_G = \frac{\alpha}{2} A_2$ and 
FR will happen if $N$ is odd at time $\tau_{FR} = \frac{\pi}{2\alpha}$. 

\medskip

It is interesting to determine if and when FR can occur in unweighted graphs. Take for this 
$\alpha = \beta = 2$. With this choice all the edges have weight 1. One then has 
$\frac{\alpha}{\beta}= \frac{1}{1}$; the condition $p$ odd is satisfied and $N$ must be even. 
In this case, $\tau_{FR} = \frac{\pi}{4}$. If $\beta = 0$, FR happens again at time 
$\tau_{FR} = \frac{\pi}{4}$, but for $N$ odd.

\section{CONCLUSION}

This paper has offered a description of the $1$-excitation dynamics of a spin chain with NNN couplings 
in terms of a quantum walk on a graph belonging to the binary Hamming scheme.

\medskip

It is known that quantum walks generated by spin chains and classical birth and death processes 
are intimately connected \cite{gvz13}. It is worth pointing out that the construction presented 
here is also related to the generalization of the Ehrenfest model developed in \cite{g09}. This reference 
offers an exact solution of a Markov process that involes nearest and next-to-nearest neighbours. 
Although this analysis in \cite{g09} is framed in terms of matrix orthogonal polynomials, the 
pentadiagonal one-step transition probability is in fact obtained, up to a constant term, as a 
quadratic expression with fixed coefficients in the Jacobi matrix of the Krawtchouk polynomials. 

\medskip

The main result of the present paper is to have obtained an example of a graph that shows FR if 
the parameters are appropriately chosen. This has been achieved via a correspondence with a spin 
chain known to possess FR. A direct graph-theoretic verification is provided in the Appendix that follows.
It would of course be of significant interest to undertake a more systematic and intrinsic study of 
FR on graphs. This has been initiated \cite{cctvz} and should be reported on in the near future.

\bigskip

\section*{ACKNOWLEDGMENTS}

A.C., C.T. and L.V. would like to thank Matthias Christandl, Gabriel Coutinho and Harmony Zhan 
for their input and support. E. L. has received a Marie Curie scholarship from the 
D\'epartement de Physique of the Universit\'e de Montr\'eal. 
The research of L.V is supported by a discovery grant from 
the National Science and Engineering Research Council (NSERC) of Canada.

%%%%%%%%%%%%%%%%%%%%%%%%%%%%%%%%%%%%%%%%%%%%%%%%%%%%%%%%%%%%%%%%%%%%%%%%%%

\bigskip

\appendix

\section{BALANCED FRACTIONAL REVIVAL ON $G_{1} \cup G_{2}$}

We have seen that the evolution governed by the graph Hamiltonian
$\overline{H}_{G} = \frac{\alpha}{2} A_{2} + \frac{\beta}{2} A_{1}$
on $G_{1} \cup G_{2}$ projects to the dynamics of single excitations
on the NNN Krawtchouk chain. 
Based on this correspondence, we have claimed that FR takes place on
$G_{1} \cup G_{2}$ owing to the properties of the spin chain.
We offer a direct graph-theoretic verification of this assertion here.

We have noted that the eigenvalues of the adjacency matrices $A_{i}$ of
the $\HH(n,2)$ scheme are given by $p_{i}(\lambda_{s}) = K_{i}(s;2,M)$
with $\lambda_{s} = M-2s$. From the recurrence relation 
(\ref{eqn:krawtchouk_recurrence}), we have in particular
\begin{equation} \label{eqn:p1}
p_{1}(\lambda_{s}) = M-2s
\end{equation}
\begin{equation} \label{eqn:p2}
p_{2}(\lambda_{s}) = \frac{1}{2}[(M-2s)^{2} - M].
\end{equation}
The Krawtchouk polynomials are also known \cite{s01} to have the following
explicit expression
\begin{equation} \label{eqn:krawtchouk_hypergeometric}
K_{n}(x;2,M) = 
	\binom{M}{n} 
    {}_{2}F_{1}\left(\begin{array}{c} -n, -x \\ -M \end{array} ; 2 \right)
\end{equation}
in terms of the hypergeometric series
\begin{equation}
{}_{2}F_{1}\left(\begin{array}{c} -n, -x \\ -M \end{array} ; 2 \right)
	= \sum_{k=0}^{\infty} \frac{(a)_{k} (b)_{k}}{(c)_{k}} \frac{x^{k}}{k!}
\end{equation}
with $(a)_{k} = a(a+1) \ldots (a+k-1)$, etc.

\par\noindent
It is readily checked from (\ref{eqn:krawtchouk_hypergeometric}) that
\begin{equation} \label{eqn:alternating}
p_{M}(\lambda_{s}) = (-1)^{s}.
\end{equation}
Denote by $E(s)$, the projectors on the eigenspaces associated to the
eigenvalues $\lambda_{s}$.
We have the spectral decomposition
\begin{eqnarray}
e^{-i\tau\overline{H}_{G}}
	& = & e^{-i\tau[\frac{\alpha}{2}A_{2} + \frac{\beta}{2}A_{1}]} \nonumber \\
    & = & \sum_{s=0}^{M} e^{-i\tau[\frac{\alpha}{2}p_{2}(\lambda_{s}) + \frac{\beta}{2}p_{1}(\lambda_{s})]} E(s).
\end{eqnarray}
Recall that $M=N-1$ in the model.
From (\ref{eqn:alternating}), we have that $A_{N-1} = \sum_{s} (-1)^{s} E(s)$.

We want to confirm that 
\begin{equation} \label{eqn:balanced_fr}
e^{-i\tau\overline{H}_{G}} 
	= e^{-i\phi'}\left(\frac{A_{0} \pm i A_{N-1}}{\sqrt{2}} \right)
\end{equation}
for the prescribed values of $\alpha$, $\beta$ and $\tau$.
Indeed the matrix relation (\ref{eqn:balanced_fr}) is tantamount to balanced
FR between antipodal vertices.

Write the eigenvalues of $\bar H_G$ as $\mathcal{E}_s$.
Using (\ref{eqn:p1}) and (\ref{eqn:p2}) one finds
\begin{eqnarray}
\tau \mathcal{E}_{s}
	& = & \tau \left[\frac{\alpha}{2}p_{2}(\lambda_{s}) + \frac{\beta}{2}p_{1}(\lambda_{s})\right] \\
    & = & \tau \left[\frac{\alpha}{4}(N-1)(N-2) + \frac{\beta}{2}(N-1)\right] \nonumber \\
	&   & \ \ \ + \ \tau\alpha s^{2} - \tau\alpha(N-1)s - \tau\beta s.
\end{eqnarray}
Let 
\begin{equation}
M_{s} = 4\tau\alpha s^{2} - 2\tau\alpha(N-1)s - 2\tau\beta s
\end{equation}
\begin{equation}
\delta = \frac{\tau}{2} [\alpha - \alpha(N-1) - \beta].
\end{equation}
Write
\begin{equation}
\phi = \tau [\frac{\alpha}{4} (N-1)(N-2) + \frac{\beta}{2}(N-1)] + \delta.
\end{equation}
For even values of $s$ we have
\begin{equation}
\tau \mathcal{E}_{2s} = \phi + M_{s} - \delta
\end{equation}
while for odd values we find
\begin{equation}
\tau\mathcal{E}_{2s+1} = \phi + M_{s} + 4\tau\alpha s + \delta.
\end{equation}
If $\beta \neq 0$, FR is supposed to occur at $\tau = \pi q/2\beta$
with $\alpha/\beta = p/q$, $p$ and $q$ coprime, $p$ odd and $q$ and $N$
of opposite parity.
Note that with these values of the parameters
\begin{equation}
M_{s} = 2\pi \left[ps^{2} - \frac{1}{2}(p(N - 1) + q)s \right].
\end{equation}
This is a multiple of $2\pi$ since the FR conditions ensure that
$p(N-1) + q$ is even. Thus $e^{-iM_{s}} = 1$.
Note also that $e^{-i(4\tau\alpha s)} = e^{-2i\pi p} = 1$.
It remains to focus on
\begin{equation}
\delta = \frac{\pi}{4} \left[p - (p(N-1) + q) \right].
\end{equation}
Given that $p(N-1) + q$ is even and $p$ is odd,
\begin{equation}
e^{i\delta} = \pm \left( \frac{1 \pm i}{\sqrt{2}} \right).
\end{equation}
For the odd values of $s$, the sign in front of $\delta$ changes so that
\begin{equation}
e^{-i\tau\mathcal{E}_{s}} = 
	e^{-i\phi'} \left( \frac{1 \pm i(-1)^{s}}{\sqrt{2}} \right)
\end{equation}
with $e^{-i\phi'} = \pm e^{-i\phi}$.

It follows that
\begin{eqnarray}
e^{-i\overline{H}_{G}} 
	& = & \sum_{s=1}^{N-1} e^{-i\tau\mathcal{E}_{s}} E(s) \\
    & = & e^{-i\phi'} \sum_{s} \frac{(1 \pm i(-1)^{s})}{\sqrt{2}} E(s) \\
    & = & e^{-i\phi'} \left( \frac{A_{0} \pm i A_{N-1}}{\sqrt{2}} \right)
\end{eqnarray}
as expected.

Similarly when $\beta = 0$, FR is predicted to happen if $N$ is odd and
$\tau = \pi/2\alpha$. In this case,
\begin{equation}
M_{s} = 2\pi s^{2} - \pi(N-1)s
\end{equation}
is a multiple of $2\pi$ because $N-1$ is even. 
Moreover $4\tau\alpha s = 2\pi s$ and $\delta = \frac{\pi}{4}N$.
This implies again (\ref{eqn:balanced_fr}).

%%%%%%%%%%%%%%%%%%%%%%%%%%%%%%%%%%%%%%%%%%%%%%%%%%%%%%%%%%%%%%%%%%%%%%%%%%


\bigskip

\section*{REFERENCES}

\begin{thebibliography}{11}

\bibitem{r04}
R.W. Robinett.
\newblock {Quantum wave packet revivals}.
\newblock {\em Physics Reports}, {\bf 392}, 1-119, 2004.

\bibitem{bcb15}
L. Banchi, E. Compagno, S. Bose.
\newblock {Perfect wave-packet splitting and reconstruction in a one-dimensional lattice}.
\newblock {\em Physical Review A}, {\bf 91}, 052323, 2015.

\bibitem{gvz16}
V. Genest, L. Vinet, A. Zhedanov.
\newblock {Quantum spin chains with fractional revival}.
\newblock {\em Annals of Physics}, {\bf 371}, 348-367, 2016.

\bibitem{b07}
S. Bose.
\newblock {Quantum communication through spin chain dynamics: an introductory review}.
\newblock {\em Contemporary Physics}, {\bf 48}, 13-30, 2007.

\bibitem{k10}
A. Kay.
\newblock {A review of perfect state transfer and its applications as a constructive tool}.
\newblock {\em International Journal of Quantum Information}, {\bf 8}, 641-676, 2010.

\bibitem{nj14}
G.M. Nikolopoulos, I. Jex.
\newblock {\em Quantum state transfer and network engineering}.
\newblock {Springer}, 2014.

\bibitem{perez-leija13}
A. Perez-Leija, R. Keil, A. Kay, H. Moya-Cessa, S. Nolte, L.-C. Kwek, B.M. Rodriguez-Lara, A. Szameit, D.N. Christodoulides.
\newblock {Coherent transport in photonic lattices}.
\newblock {\em Physical Review A}, {\bf 87}, 012309, 2013.

\bibitem{chapman16}
R.J. Chapman, M. Santandrea, Z. Huang, G. Corrielli, A. Crespi, M.-H. Yung, R. Osellame, A. Peruzzo.
\newblock {Experimental perfect state transfer of an entangled photonic qubit}.
\newblock {\em Nature Communications}, {\bf 7}, 11335, 2016.

\bibitem{cddekl05}
M. Christandl, N. Datta, T.C. Dorlas, A. Ekert, A. Kay, A.J. Landahl.
\newblock {Perfect transfer of arbitrary states in quantum spin networks}.
\newblock {\em Physical Review A}, {\bf 71}, 032312, 2005.

\bibitem{acde04}
C. Albanese, M. Christandl, N. Datta, A. Ekert.
\newblock {Mirror inversion of quantum states in linear registers}.
\newblock {\em Physical Review Letters}, {\bf 93}, 230502, 2004.

\bibitem{g12}
C. Godsil.
\newblock {State transfer on graphs}.
\newblock {\em Discrete Mathematics}, {\bf 312}, 123-147, 2012.

\bibitem{kt11}
V. Kendon, C. Tamon.
\newblock {Perfect state transfer in quantum walks on graphs}.
\newblock {\em Journal of Computational and Theoretical Nanoscience}, {\bf 8}, 422-433, 2011.

\bibitem{cfg02}
A. Childs, E. Farhi, S. Gutmann.
\newblock {An example of the difference between quantum and classical random walks}.
\newblock {\em Quantum Information Processing}, {\bf 1}, 35-43, 2002.

\bibitem{cvz17}
M. Christandl, L. Vinet, A. Zhedanov.
\newblock {Analytic next-to-nearest neighbor XX models with perfect state transfer and fractional revival}.
\newblock {\em Physical Review A}, {\bf 96}(3), 032335, 2017.

\bibitem{vz12}
L. Vinet, A. Zhedanov.
\newblock {How to construct spin chains with perfect state transfer}.
\newblock {\em Physical Review A}, {\bf 85}, 012323, 2012.

\bibitem{bcn89}
A.E. Brouwer, A.M. Cohen, A. Neumaier.
\newblock {\em Distance-Regular Graphs}.
\newblock Springer, 1989.

\bibitem{bi84}
E. Bannai, T. Ito.
\newblock {\em Algebraic Combinatorics}.
\newblock Benjamin/Cummings, 1984.

\bibitem{b90}
E. Bannai.
\newblock {Orthogonal polynomials in coding theory and algebraic combinatorics}.
\newblock In {\em Orthogonal Polynomials: Theory and Practice}, P. Nevai (ed.), NATO ASI series, Vol. 294, 25-53, Springer, 1990.

\bibitem{s01}
D. Stanton.
\newblock {Orthogonal polynomials and combinatorics}.
\newblock In {\em Special Functions 2000: Current perspective and future directions}, J. Boustoz, M.E.H. Ismail, S. Suslov (eds.), NATO science series, Vol. 30, 389-409, Springer, 2001.

\bibitem{cs90}
L. Chihara, D. Stanton.
\newblock {Zeros of generalized Krawtchouk polynomials}.
\newblock {\em Journal of Approximation Theory}, {\bf 60}, 43-57, 1990.

\bibitem{gvz13}
F.A. Gr\"{u}nbaum, L. Vinet, A. Zhedanov.
\newblock {Birth and death processes and quantum spin chains}.
\newblock {\em Journal of Mathematical Physics}, {\bf 54}, 062101, 2013.

\bibitem{g09}
F.A. Gr\"{u}nbaum.
\newblock {Block tridiagonal matrices and a beefed-up version of the Ehrenfest urn model}.
\newblock In {\em Modern Analysis and Applications}, V.M. Adamyan et al. (eds.), Operator Theory: Advances and Applications, Vol. 190, 267-277, Springer, 2009.

\bibitem{cctvz}
A. Chan, G. Coutinho, C. Tamon, L. Vinet, H. Zhan.
\newblock In preparation.

\end{thebibliography}
\end{document}